\documentclass[nofootinbib,a4paper,aps,prd,10pt,superscriptaddress,showkeys]{revtex4}

\usepackage{graphicx}
\usepackage{amsfonts}
\usepackage{amssymb} 
\usepackage{amsmath}
\usepackage{hyperref}
\usepackage{natbib}
\usepackage{float}
\usepackage{dcolumn}
\usepackage{bm}
\usepackage{latexsym,color}

\def\mnras{Mon. Not. Roy. Astr. Soc.}

\newcommand{\bea}{\begin{eqnarray}}
\newcommand{\eea}{\end{eqnarray}}
\newcommand{\be}{\begin{equation}}
\newcommand{\ee}{\end{equation}}

\begin{document}

\title{Adiabatic theory of motion of bodies in the Hartle-Thorne spacetime}

\author{Gulnara~\surname{Sulieva}}
\email[]{sulieva.gulnara0899@gmail.com}
\affiliation{%
Al-Farabi Kazakh National University, Al-Farabi av. 71, 050040 Almaty, Kazakhstan.
}

\author{Kuantay~\surname{Boshkayev}}
\email[]{kuantay@mail.ru}
\affiliation{%
Al-Farabi Kazakh National University, Al-Farabi av. 71, 050040 Almaty, Kazakhstan.
}
\affiliation{National Nanotechnology Open Laboratory,  Almaty 050040, Kazakhstan.}

\author{Gulmira~\surname{Nurbakyt}}
\email[]{gumi-nur@mail.ru}
\affiliation{%
Al-Farabi Kazakh National University, Al-Farabi av. 71, 050040 Almaty, Kazakhstan.
}

\author{Hernando~\surname{Quevedo}}
\email[]{quevedo@nucleares.unam.mx}
\affiliation{%
Al-Farabi Kazakh National University, Al-Farabi av. 71, 050040 Almaty, Kazakhstan.
}
\affiliation{%
Instituto de Ciencias Nucleares, Universidad Nacional Aut\`onoma de M\`exico, Mexico. 
}
\affiliation{%
Dipartimento di Fisica and ICRA, Universit\`a di Roma “La Sapienza”, Roma, Italy.
}

\author{Aliya~\surname{Taukenova}}
\email[]{aliya\_tauken@mail.ru}
\affiliation{%
Al-Farabi Kazakh National University, Al-Farabi av. 71, 050040 Almaty, Kazakhstan.
}
\affiliation{National Nanotechnology Open Laboratory,  Almaty 050040, Kazakhstan.}

\author{Abylaikhan~\surname{Tlemissov}}
\email[]{tlemissov-ozzy@mail.ru}
\affiliation{%
Institute of Physics, Silesian University in Opava, Bezrucovo nam. 13, CZ-74601 Opava, Czech Republic.
}

\author{Zhanerke~\surname{Tlemissova}}
\email[]{kalymova.erke@mail.ru}
\affiliation{%
Institute of Physics, Silesian University in Opava, Bezrucovo nam. 13, CZ-74601 Opava, Czech Republic.
}

\author{Ainur~\surname{Urazalina}}
\email[]{y.a.a.707@mail.ru}
\affiliation{%
Al-Farabi Kazakh National University, Al-Farabi av. 71, 050040 Almaty, Kazakhstan.
}
\affiliation{National Nanotechnology Open Laboratory,  Almaty 050040, Kazakhstan.}

\date{\today}

\begin{abstract}
We study the motion of test particles in the gravitational field of a rotating and deformed object within the framework of the adiabatic theory. For this purpose, the Hartle-Thorne metric written in harmonic coordinates is employed in the post-Newtonian approximation where the adiabatic theory is valid. As a result, we obtain the perihelion shift formula for test particles orbiting on the equatorial plane of a rotating and deformed object. Based on the perihelion shift expression, we show that the principle of superposition is valid for the individual effects of the gravitational source mass, angular momentum and quadrupole moment. The resulting formula was applied to the inner planets of the Solar system. The outcomes are in a good agreement with observational data. It was also shown that the corrections related to the Sun's angular moment and quadrupole moment have little impact on the perihelion shift. On the whole, it was demonstrated that the adiabatic theory, along with its simplicity, leads to correct results, which in the limiting cases correspond to the ones reported in the literature.
\end{abstract}


\maketitle

\section{Introduction}
In most cases, real astrophysical objects rotate and their shapes are different from a sphere. Therefore, when one considers the motion of test particles in the gravitational field of real objects, it is necessary to account for the influence of both proper rotation and deformation of the source.
A convenient way to consider the geometry of the source is to study its multipole moments of which the most important are the mass $M$, angular momentum $J$, and quadrupole moment $Q$. The solution to the field equations for a static, spherically symmetric object in vacuum is well-known in the literature as the Schwarzschild metric \cite{1916AbhKP1916..189S}. This solution describes new effects that could not be explained within the classical Newtonian theory of gravity \cite{1973grav.book.....M}, \cite{2013grsp.book.....O}. In 1918, Lense and Thirring derived an approximate external solution that takes into account the rotation of the source up to the first order in the angular momentum \cite{1918PhyZ...19..156L}. According to this work, rotation generates and additional gravitational field which leads to the dragging of inertial frames (known as the Lense-Thirring effect). In 1959, Erez and Rosen derived a solution for a static, axially symmetric object by including of a quadrupole parameter \cite{1959}. However, the first approximate solution that takes into account both angular momentum and quadrupole moment was found by Hartle and Thorne in 1968 \cite{1967ApJ...150.1005H,1968ApJ...153..807H}. This solution allows us to investigate the external gravitational field of astrophysical objects, starting from massive main sequence stars up to neutron and quark stars \cite{2004MNRAS.350.1416B}.
It should be mentioned that there are several vacuum exact solutions to the Einstein field equation, which account for higher-order  multipole moments with additional parameters  such as electric charge, dilatonic charge, scalar fields, etc \cite{2015CQGra..32p5010A,2017EPJC...77..180A,2020JPhCS1690a2143B, 2021EPJC...81..475M}. However, for simplicity, here we will focus on the approximate Hartle and Thorne solution and will study the motion of test bodies within  the adiabatic theory.

An interesting approach for studying the motion of test particles in general relativity was proposed by Abdildin \cite{1988mtge.book.....A}, \cite{2006mtge.book.....A}, by using the conceptual framework developed by Fock \cite{1961tstg.book.....A}. In Ref.~\cite{1988mtge.book.....A}, the Fock metric was  generalized to consider  the rotation of the source (up to  the second order in the angular momentum) and its internal structure in the post-Newtonian ($\sim 1/c^2$) approximation, where $c$ is the speed of light in vacuum. This extended Fock metric was originally presented  in harmonic coordinates, 
which facilitate the study of the motion of test particles by using the vectors associated to the trajectories.  
One of the most important consequences of Abdildin's works was the implementation  of the adiabatic theory to study the motion of bodies in general relativity \cite{2006mtge.book.....A}, which drastically simplifies the form of the equations of motion derived previously in  \cite{1957RvMP...29..398I,1962RvMP...29..398I}.
In this work, we will show this advantage explicitly for the motion of test particles in the gravitational field of a rotating deformed object.

The work is organized as follows. In Section \ref{sec:adtheor}, we introduce the basic concepts of the adiabatic theory. In Section \ref{sec:ht}, we present the external Hartle-Thorne solution, which is then implemented  in Section \ref{sec:adtheorht} within the framework of the  adiabatic theory to obtain an expression for the perihelion shift. Then, in Section \ref{sec:res}, we compute the shift  for the inner planets of the Solar system. Finally, Section \ref{sec:con} contains the conclusions of our analysis. 


\section{Adiabatic theory}\label{sec:adtheor}
The application of adiabatic theory for the investigation of motion in general relativity, as proposed in \cite{2006mtge.book.....A} for closed orbits, is based on the use of the vector elements of the orbits,  asymptotic methods of the theory of nonlinear oscillations, and adiabatic invariants.  

The main idea is that the motion can be described by a Lagrangian which is essentially the perturbation of a known Lagrangian. Consider, for instance, the Kepler problem for the motion of a relativistic particle in a central field. Then, corresponding perturbed Lagrangian function can be expressed as 
\be
L=-mc^2+\frac{mv^2}{2}+\frac{Gmm_0}{r}+F(\vec{r},\vec{v}),
\ee
where $F$ is the perturbation function. Accordingly, the corresponding Hamilton function is written as
\be
H=mc^2-\frac{p^2}{2m}-\frac{Gmm_0}{r}-F(\vec{r},\vec{p}),
\ee
where $\vec{p}=\partial L/\partial \vec{v}$ is the momentum of the test particle.

The motion of a test particle can be described by the  the orbital angular momentum vector $\vec{M}$ and the Laplace-Runge-Lenz vector $\vec{A}$, which are integrals of motion defined as:
\bea
\vec{M}&=&\left[\vec{r} \times \vec{p} \right], \\
\vec{A}&=&\left[\frac{\vec{p}}{m} \times \vec{M} \right]-\frac{Gm_0 m}{r}\vec{r} , \quad A=Gm_0 me ,
\eea
where $A$ is the magnitude (absolute value) of the Laplace-Runge-Lenz vector, $\vec{r}$ is the radius vector of the test particle, $G$ is the gravitational constant, $m_0$ is the mass of a gravitational source (central object), $m$ is the mass of the test particle, and $e$ is the orbit eccentricity. 
The vectors $\vec{M}$ and $\vec{A}$  characterize the shape and position of the orbit in space. Namely, the vector $\vec{M}$ is directed perpendicularly to the orbit plane and the vector $\vec{A}$ is directed towards the perihelion of the orbit. Thus, one can write the equations of motion in a general form as follows:
\bea\label{eq:MA}
\frac{d\vec{M}}{dt}&=&\frac{dM}{dt}\vec{e}_M+\left[\vec{\Omega}\times \vec{M} \right], \label{eq:M}\\
\frac{d\vec{A}}{dt}&=&\frac{dA}{dt}\vec{e}_A+\left[\vec{\Omega}\times \vec{A} \right], \label{eq:A}
\eea
where $\vec{e}_M$, $\vec{e}_A$ are the unit vectors directed along $\vec{M}$ and $\vec{A}$, respectively, and $\vec{\Omega}$ is the angular velocity of rotation of the ellipse ``as a whole'', which is the sought function in this theory. The explicit form of $\vec{\Omega}$ depends on the considered physical system. In Ref.~\cite{1988mtge.book.....A}, it is shown that 
the angular velocity can be computed as
\be\label{eq:omega}
\vec{\Omega}=\frac{\partial \overline{H}}{\partial \vec{M}},
\ee
where  $\overline{H}$ is the Hamiltonian averaged over the period of the test particle's Keplerian orbit. The averaged Hamiltonian depends on the orbital angular momentum $\vec{M}$ and the adiabatic invariant $M_0$ of the system
\be\label{eq:adinv}
M_0 =\frac{M}{\sqrt{1-A^2/\alpha^2}},
\ee
where $\alpha=Gmm_0$.

The knowledge of the angular velocity $\vec{\Omega}$ allows us to investigate many relativistic effects without solving Eqs.~\eqref{eq:M} and \eqref{eq:A} explicitly. The  invariant Eq.~\eqref{eq:adinv} allows to write Eqs.~\eqref{eq:M} an \eqref{eq:A} in a more compact form as
\bea\label{eq:MA2}
\frac{d\vec{M}}{dt}&=&\frac{dM}{dt}\vec{e}_M+\left[\vec{\Omega}\times \vec{M} \right], \label{eq:M2}\\
\frac{d\vec{e}_A}{dt}&=&\left[\vec{\Omega}\times \vec{e}_A \right]. \label{eq:A2}
\eea

Thus, in the adiabatic theory, Eqs.~\eqref{eq:M2} and \eqref{eq:A2} and the expression \eqref{eq:omega} are the mathematical basis for the investigation of the motion of bodies. In other words, these equations completely solve the problem of evolution in the quasi-Kepler problem.

\begin{figure}[ht]\label{fig:pic}
\includegraphics[width=0.5\linewidth]{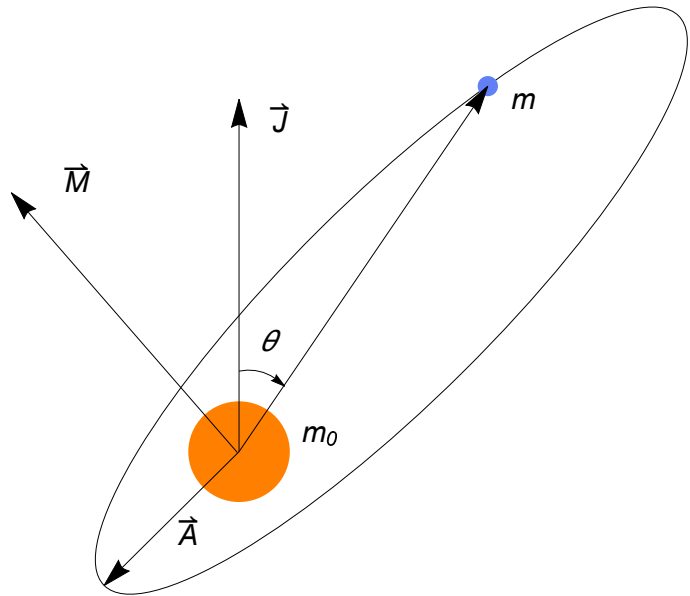} 
\caption{Schematic illustration of a central object and a test particle with its vector elements, where $\theta$ is the polar angle between the $z$ axis and the radius vector $\vec{r}$}
\end{figure}

In Fig.~\ref{fig:pic}, we show the position of the vector elements and the proper angular momentum of the central object $\vec{J}$, which is directed along the $z$ axis. Note that when $\theta=\pi/2$ the directions of $\vec{M}$ and $\vec{J}$ coincide with the $z$ axis. 


\section{The Hartle-Thorne metric}\label{sec:ht}
The Hartle-Thorne metric is an approximate vacuum solution of the Einstein field equations. It describes well enough the gravitational field of rotating deformed astrophysical objects and, therefore, it is chosen as an example in this work. Its general form (in geometric units $G=c=1$) in spherical coordinates $(t, R, \Theta, \phi)$ is given by
\bea \label{eq:metr}
ds^2&=&-\left(1-\frac{2m_0}{R}\right) \left[1+2k_1 P_2(\cos\Theta)-2\left(1-\frac{2m_0}{R} \right)^{-1} \frac{J^2}{R^4}(2\cos^2 \Theta -1) \right] dt^2\nonumber
\\&+&\left(1-\frac{2m_0}{R} \right)^{-1}\left[1-2\left(k_1 -\frac{6J^2}{R^4} \right) P_2(\cos\Theta)-2 \left(1-\frac{2m_0}{R} \right)^{-1} \frac{J^2}{R^4}\right]dR^2\nonumber
\\&+&R^2[1-2k_2 P_2(\cos\Theta)](d\Theta^2 +\sin^2 \Theta d\phi^2)-\frac{4J}{R}\sin^2 \Theta dt d\phi,
\eea
where
\bea
k_1 &=&\frac{J^2}{m_0R^3}\left(1+\frac{m_0}{R} \right) +\frac{5}{8}\frac{Q-J^2 /m_0}{m_0^3}Q_2^2 (x),\\
k_2 &=&k_1 +\frac{J^2}{R^4}+\frac{5}{4}\frac{Q-J^2 /m_0}{m_0^2 R}\left(1-\frac{2m_0}{R} \right)^{-1/2} Q_2^1 (x),
\eea
are functions of the $R$ coordinate, and
\bea
Q_2^1(x)&=&(x^2 -1)^{1/2}\left[\frac{3x}{2}\ln\left(\frac{x+1}{x-1} \right) -\frac{3x^2 -2}{x^2 -1} \right] , \nonumber\\
Q_2^2(x)&=&(x^2-1)\left[\frac{3}{2}\ln\left(\frac{x+1}{x-1} \right) -\frac{3x^3 -5x}{(x^2 -1)^2} \right],
\eea
are the associated Legendre functions of the second kind 
\cite{1977eomph...29..398I,1972hmfw.book.....A}, $P_2(\cos\Theta)$ is the Legendre polynomial, and $x=R/m_0-1$. This metric is characterized by three parameters: the source mass $m_0$, angular momentum $J$ (up to the second order), and quadrupole moment $Q$ (up to the first order). 

The Hartle-Thorne metric describes the gravitational field of slowly rotating and slightly deformed astrophysical objects \cite{2003LRR.....6....3S}. The metric \eqref{eq:metr} can be reduced by appropriate coordinate transformations to the Fock metric \cite{2013CFHTM.....6....3S}, to the Kerr metric \cite{2015CKHTM.....6....3S}, and to the Erez-Rosen metric \cite{2019htvsers.....6....3S,2020Chters.....6....3S} in the corresponding limiting cases.
For the purpose of this work, the metric \eqref{eq:metr} must be written in harmonic coordinates and expanded in a series of powers of $1/c^2$. 

Harmonic coordinates are important for many problems in general relativity \cite{1961tstg.book.....A}. Such coordinates are associated with the conditions under which spacetime is considered homogeneous and isotropic at large distances from the gravitational field source. In turn, a consequence of the homogeneity and isotropy of the spacetime is the conservation of energy, momentum and angular momentum, which are in fact first integrals of the motion equations. In general, harmonic coordinates can be used in the study of gravitational fields generated by ordinary stars \cite{1973PhT....26f..57W}, black holes \cite{1998liu....26f..57W}, as well as in the study of quantum gravity \cite{2018Univ....4..103G}, supergravity \cite{1994AnPhy.230..201G}, and in numerical relativity \cite{2002PhRvD..65d4029G}.

It should be emphasized that the geodesics in the Hartle-Thorne spacetime have been studied in the literature both analytically and numerically \cite{2003gr.qc....12070A, 2013NCimC..36S..31B,2016IJMPA..3141006B}. Here, unlike in the literature, we employ an alternative method to derive the perihelion shift formula in  post-Newtonian physics. 

\section{The method}\label{sec:adtheorht}
As already mentioned, in the present work we need the Hartle-Thorne metric expanded in powers of $1/c^2$. In harmonic coordinates it is written as follows \cite{2013CFHTM.....6....3S,2012PhRvD..86f4043B}:
\bea \label{eq:metrharm}
ds^2 &=& \left[1-\frac{2Gm_0}{c^2 r}+\frac{2GQ}{c^2 r^3}P_2(\cos\theta)+\frac{2G^2 m_0^2}{c^4 r^2}-\frac{4G^2 m_0 Q}{c^4 r^4}P_2(\cos\theta)\right]c^2 dt^2\nonumber
\\&-& \left[1+\frac{2Gm_0}{c^2 r}-\frac{2GQ}{c^2 r^3}P_2(\cos\theta)\right][dr^2 +r^2 (d\theta^2 +\sin^2\theta d\phi^2)]+\frac{4GJ}{c^2 r}\sin^2\theta dtd\phi\ .
\eea
This representation allows us to explicitly identify relativistic corrections. Thus, in the $g_{tt}$ component of the metric tensor, the first three terms refer to the Newtonian theory and the last two terms to the relativistic theory because of the multiplier $c^2$ outside the parenthesis. Moreover, terms proportional to  $ 1/c^2$ also appear in the spatial part of the metric.
	
Now, directly from the metric \eqref{eq:metrharm} one finds the Lagrange function of the test particle
\bea \label{eq:lagr}
L&=&-mc\frac{ds}{dt}=-mc^2 +\frac{mv^2}{2}+\frac{Gmm_0}{r}-\frac{GmQ}{r^3}P_2(\cos\theta)\nonumber
\\&+&\frac{m}{2c^2}\left[\frac{v^4}{4}+\frac{3Gm_0 v^2}{r}-\frac{G^2 m_0^2}{r^2}-\frac{3Gv^2 Q}{r^3}P_2(\cos\theta)+\frac{2G^2 m_0 Q}{r^4}P_2(\cos\theta)-\frac{4G \left(\vec{v}\cdot\left[\vec{r}\times\vec{J}\right]\right)}{ r^3}\right],
\eea
and besides
\be
\vec{v}=\frac{d\vec{r}}{dt}, \quad v^2 =\frac{dr^2 +r^2 (d\theta^2 +\sin\theta^2 d\phi^2)}{dt^2}.
\ee
Only in harmonic and isotropic coordinates, it is possible to write the linear velocity in the form indicated above.

Next, it is necessary to derive the Hamiltonian, which we will subsequently average. The expression to determine the Hamilton function is given as \cite{1969mech.book.....L}:
\be
H=(\vec{p}\cdot\vec{v})-L.
\ee
First, we look for the form of the generalized momentum $\vec{p}$. Thus,
\be \label{eq:genmom}
\vec{p}=\frac{\partial L}{\partial \vec{v}}=\left[1+\frac{v^2}{2c^2}+\frac{3Gm_0}{c^2 r}-\frac{3GQ}{c^2 r^3}P_2(\cos\theta)\right]m\vec{v}-\frac{2Gm}{c^2 r^3}\left[\vec{r}\times\vec{J}\right].
\ee
Taking into account \eqref{eq:lagr} - \eqref{eq:genmom}, the Hamiltonian takes the following form:
\bea \label{eq:hamilt}
H&=&mc^2 +\frac{p^2}{2m}-\frac{Gm_0 m}{r}+\frac{GmQ}{r^3}P_2(\cos\theta)-\frac{p^4}{8c^2 m^3}-\frac{3Gm_0 p^2}{2c^2 mr}\nonumber
\\&+&\frac{G^2 m_0^2 m}{2c^2 r^2}+\frac{3GQp^2}{2c^2 mr^3}P_2(\cos\theta)-\frac{G^2 m_0 mQ}{c^2 r^4}P_2(\cos\theta)+\frac{2G \left(\vec{p}\cdot\left[\vec{r}\times\vec{J}\right]\right)}{c^2 r^3}.
\eea
For simplicity, we consider the motion of test particle on the equatorial plane, i.e., $\theta = \pi/2$. Now, according to the adiabatic theory, we should average each term in \eqref{eq:hamilt} over the period $T$, where the average of a function $f$ is defined as:
\be \label{eq:aver}
\overline{f}=\frac{1}{T}\underset{0}{\overset{T}{\int }}fdt.
\ee
In this work, for convenience, averaging is carried out using the non-relativistic orbital angular momentum $M$  in  polar coordinates
\be
M=mr^2 \frac{d\phi}{dt},
\ee
which allows us to change from an integral  over $t$ to and integral over $\phi$. Here, we use the solution to the Kepler problem \cite{1969mech.book.....L}
\be
r=\frac{P}{1+e\cos \phi}, \quad 0<\phi<2\pi,
\ee
where $e$ is the orbit eccentricity as before, $P$ is the semilactus rectum, and $\phi$ is the polar angle. Therefore, it turns out that
\be
\overline{f}=\frac{1}{T}\underset{0}{\overset{2\pi}{\int }}f(\phi)\frac{dt}{d\phi}d\phi=\frac{m}{TM}\underset{0}{\overset{2\pi}{\int }}f(\phi)r^2 d\phi.
\ee
In addition, to average terms in Eq.~\eqref{eq:hamilt} with the momentum $\vec{p}=m\vec{v}$, we use the following form of the test particle velocity:
\be
\vec{v}=\frac{M}{mP}\left\{-\vec{i}\sin \phi +\vec{j}(e+\cos \phi)\right\}.
\ee

It is also important to mention that one is free to choose the direction of the central body rotation. For simplicity and practical purposes, it is preferred to align it along the $z$ axis as $\vec{J}=J\vec{k}$. For a test particle moving in the equatorial plane, its orbital angular momentum direction coincides with the proper angular momentum of the central body, i.e.,  $\vec{M} \uparrow \uparrow \vec{J}$, hence $\vec{M}=M\vec{k}$.
 
Applying Eq.~\eqref{eq:aver} to each term in Eq.~\eqref{eq:hamilt} and using the formula for the period, $T=2\pi M_0^3/m\alpha^2$ \cite{1969mech.book.....L}, one obtains the averaged Hamilton function:
\bea
\overline{H}&=&mc^2-\frac{m\alpha^2}{2M_0^2}-\frac{3m\alpha^4}{c^2 M_0^3 M}+\frac{15m\alpha^4}{8c^2 M_0^4}+\frac{2m^2\alpha^4 J}{m_0 c^2 M_0^3 M^2} \nonumber
\\&-&\frac{m^3 \alpha^4 Q}{2m_0 M_0^3 M^3}-\frac{3m^3 \alpha^6 Q}{2m_0 c^2 M_0^3 M^5}+\frac{5m^3 \alpha^6Q}{4m_0 c^2 M_0^5 M^3}.
\eea
As expected, the averaged Hamiltonian depends on the adiabatic invariant $M_0$ and the orbital angular momentum $M$.
 
The next step is to find the form of the angular velocity $\vec{\Omega}$. For this, according to Eq.~\eqref{eq:omega}, we need to take the partial derivative of $\overline{H}$ with respect to $\vec{M}$. The result is the following:
\be
\vec{\Omega}=\left(\frac{3m\alpha^4}{c^2 M_0^3 M^2}-\frac{4m^2 \alpha^4 J}{m_0 c^2 M_0^3 M^3}+\frac{3m^3 \alpha^4 Q}{2m_0 M_0^3 M^4}+\frac{15m^3 \alpha^6 Q}{2m_0 c^2 M_0^3 M^6}-\frac{15m^3 \alpha^6 Q}{4m_0 c^2 M_0^5 M^4}\right)\vec{e}_M.
\ee
Finally, to find the perihelion shift angle $\Delta g$, we multiply the angular velocity module $\vec{\Omega}$ by the orbital period $T$ of a test particle. Thereby, we get the form:
\be\label{eq:deltag}
\Delta g=\frac{6\pi Gm_0}{c^2 P}-\frac{8 \pi GmJ}{c^2 MP}+\frac{3\pi Q}{m_0 P^2}+\frac{15\pi GQ (1+e^2)}{2c^2 P^3},
\ee
where $P=M^2 /m\alpha=a(1-e^2)$, $a$ is the semi-major axis of the orbit.
	
From Eq.~\eqref{eq:deltag}, we can see that for  the considered problem the principle of superposition of effects is valid due to the approximate character of the solution as given in terms of the source mass, angular momentum and quadrupole moment. The first term corresponds to the solution of the Schwarzschild problem (i.e., due to the curvature of spacetime caused by the mass of the central body); the second term arises as a result of accounting for the rotation of the source (it appears as the frame dragging effect - the Lense-Thirring effect); the third term is the classical correction due to the quadrupole moment, as a consequence of the source deformation; and the fourth term is the relativistic correction for the quadrupole moment.

It should be noted, that the effect of perihelion shift (rotation) in the Schwarzschild problem is associated with the appearance in the Hamiltonian of the dependence on orbital momentum $M$. In classical mechanics, i.e., in the Kepler problem, there is no such dependence and the perihelion remains motionless.
	
Furthermore, the resulting expression \eqref{eq:deltag} for the perihelion shift in the limits
\begin{itemize}
\item $J=0$ and $Q=0$ reduces to the Schwarzschild case \cite{2006mtge.book.....A}, \cite{1975ctf..book.....L};
\item $J \neq 0$ and $Q=0$  reduces to the Lense-Thirring effect \cite{2006mtge.book.....A}, \cite{1975ctf..book.....L};
\item $J=0$ and $Q \neq 0$ reduces to the case of a static deformed source \cite{2018RecContrtoPhys..67};
\item $J\neq 0$ and $Q \neq 0$ reduces to the case of the extended Fock metric \cite{2012PhRvD..86f4043B}.
\end{itemize} 

To be more precise, in the extended Fock metric $Q=\kappa J^2/(m_0 c^2)$, different values of $\kappa$ correspond to the following limiting cases (in the $ 1/c^2$ approximation):  
	\begin{itemize}
		\item $\kappa=1$ for the Kerr metric;
		\item $\kappa=4/7$ for the liquid body metric;
		\item $\kappa=15/28$ for the solid body metric.
	\end{itemize}

When comparing, one must keep in mind that in Ref.~\cite{2006mtge.book.....A} the angular momentum of the central body is denoted by $S_0=J$ and quadrupole moment in \cite{2018RecContrtoPhys..67} is denoted by $D$, which is linked with $Q$ of this work by $Q=-D/2$.

	
\section{Analysis of the results}\label{sec:res}
Now we apply Eq.~\eqref{eq:deltag} to estimate the perihelion shift of the Solar system inner planets: Mercury, Venus and Earth. For calculations, we use the Sun mass, radius, angular momentum and quadrupole moment. The test body is a planet so that its shape and size are not taken into account. Usually, the quadrupole parameter $J_2$ is chosen instead of the quadrupole moment $Q$. There is a straightforward relation between them \cite{1975ctf..book.....L}
\be
J_2=\frac{Q}{4m_0 R^2},
\ee
where $m_0$, $R$ are the Sun mass and radius, correspondingly. The last experimentally measured value of the solar quadrupole parameter is given in \cite{2017AsJ.121} as $J_2=(2.25 \pm 0.09) \cdot 10^{-7}$. As for the Sun angular moment, unfortunately, there are no values in the literature based on observational and experimentally studied data. Therefore, to find it, we can use the general formula for the angular momentum \cite{1969mech.book.....L}:
\be
J=I\omega,
\ee
where $\omega$ is the angular velocity of a body rotating around its axis and $I=\frac{2}{5} m_0 R^2$ is the moment of inertia of a sphere. It should be noted that the rotation of the Sun is differential, i.e., it decreases with the distance from the equator to the poles. However, as an example, one can choose the value of the angular velocity on the equator $\omega=2.9 \cdot 10^{-6}$ rad/s \cite{1994ssevol..book.....K}. So, the Sun angular momentum is approximately $J=2.79 \cdot 10^ {42}$  kg$\cdot$m$^2$/s.

Table~\ref{tabular: tab1} presents the orbital parameters of Mercury, Venus, and the Earth \cite{1993tandexpGP..book.....W}, \cite{2006LRinRel.100}. Moreover, all the corrections given in Eq.~\eqref{eq:deltag} are calculated separately to estimate the individual contribution of each effect. All values are calculated for 100 Earth years.
\begin{table}[H]
 	\caption{Orbital parameters and perihelion shift angles of Mercury, Venus, and the Earth}
 	\label{tabular: tab1}
 	\begin{center}
 		\begin{tabular}{|c|c|c|c|}
 			\hline
 			Planets &  Mercury &  Venus & Earth \\
 			\hline
 			Semi-major axis, $a$ (km) & 57909082 & 108208600 & 149597870 \\ 
 			\hline
 			Eccentricity, $e$ & 0.2056 & 0.0068 & 0.0167 \\ 
 			\hline
 			Semilactus rectum, $P$ (km) & 55460308 & 108203681 & 149556105 \\ 
 			\hline
 			Sidereal period, $T$, (earth days) & 87.968 & 224.695 & 365.242 \\ 
 			\hline
 			$6\pi Gm_0/c^2 P$ & 43'' & 8.63'' & 3.84'' \\ 
 			\hline
 			$8 \pi GmJ/c^2 MP$ & 0.116'' & 0.017'' & 0.006'' \\ 
 			\hline
 			$3\pi Q/m_0 P^2$ &0.03'' & 0.003'' & 0.001'' \\ 
 			\hline
 			Observational data &(43.11$\pm$0.45)''&(8.4$\pm$4.8)''&(5.0$\pm$1.2)''\\ 
 			\hline
 				\end{tabular}
 			\end{center}
 	\end{table}

As can be seen from Table~\ref{tabular: tab1}, the Mercury orbit has the largest value of the perihelion shift. This is due to several factors. Firstly, Mercury is closer than other planets to the Sun and, therefore, is more influenced by its gravitational field. Secondly, Mercury rotates  around the Sun faster (in one hundred Earth years, it makes about 415 revolutions, while Venus makes about 162 revolutions, only). 

As for Mercury, Venus and the Earth, a significant contribution to the perihelion shift is made by the effect related to the Sun mass. Compared to this, the correction due to the Sun rotation for all three planets has less of an impact; the classical quadrupole moment correction is even less than the latter. In this case, the relativistic quadrupole moment correction $15\pi GQ (1+e^2)/(2c^2 P^3)$ is negligible in magnitude, so its contribution can be ignored for the Solar system. 

The calculated values are in good agreement with the observational data. According to observations, the measurement error for Mercury is 0.45'', for Venus is 4.8'', and for the Earth is 1.2''. This is due to the fact that the perihelion shift is more certain for orbits with a large eccentricity (as for Mercury). If the orbit is close to circular in shape (as for Venus), it becomes much more difficult to observe the displacement of its perihelion.


\section{Conclusion}\label{sec:con}
In this article, we considered the motion of test particles in the gravitational field of a slowly rotating and slightly deformed object within the framework of the adiabatic theory. For this purpose, the Hartle-Thorne metric was used, expanded in a series in powers of $1/c^2$, and written in harmonic coordinates.

The perihelion shift expression was derived for the Hartle-Thorne metric. The influence of the central body rotation and deformation on the test particles trajectory was shown. It was also demonstrated that the resulting formula satisfies the principle of superposition of relativistic effects due to the approximate character of the solution as given in terms of the source mass, angular momentum and quadrupole moment. In the limiting cases, the perihelion shift formula corresponds to the values presented in literature.

As an example, the results of this work were applied to the inner planets of the Solar system. As expected, the main influence on the planets motion is exerted by the curvature of spacetime related to the Sun mass. Although taking into account the Sun rotation and deformation has a minor role, the obtained formula for the perihelion shift can be applied to exoplanetary or other relativistic systems, where their contribution may be more significant.

It would also be interesting to study the motion of test particles in the non-equatorial plane applying both perturbation and adiabatic theories. This task will be considered in future studies.

\begin{acknowledgments}
KB, AU and AT acknowledge the Ministry of Education and Science of the Republic of Kazakhstan, Grant: IRN AP08052311.
\end{acknowledgments}

\end{document}